\documentclass[twocolumn,aps,prl,superscriptaddress]{revtex4-1}
\usepackage{amsmath}
\usepackage{amssymb}
\usepackage{color}
\usepackage{graphicx}
\usepackage{times}
\usepackage{ulem}
\usepackage[breaklinks]{hyperref}
\hypersetup{
colorlinks=true,
linkcolor=blue,
citecolor=blue,
filecolor=blue,
urlcolor=blue}

\newcommand{\be}{\begin{equation}}
\newcommand{\ee}{\end{equation}}
\newcommand{\nn}{\nonumber\\}

\newcommand{\p}{\partial}
\newcommand{\la}{\langle}
\newcommand{\ra}{\rangle}

\renewcommand{\vec}[1]{{\bf #1}}

\begin{document}
\title{Geometric photon-drag effect and nonlinear shift current in centrosymmetric crystals}

\author{Li-kun Shi}
\affiliation{Division of Physics and Applied Physics, Nanyang Technological University, Singapore 637371}

\author{Dong Zhang}
\affiliation{SKLSM, Institute of Semiconductors, Chinese Academy of Sciences, P.O. Box 912, Beijing 100083, China}
\affiliation{CAS Center for Excellence in Topological Quantum Computation,
University of Chinese Academy of Sciences, Beijing 100190, China}
\affiliation{Beijing Academy of Quantum Information Sciences, Beijing 100193, China}

\author{Kai Chang}
\affiliation{SKLSM, Institute of Semiconductors, Chinese Academy of Sciences, P.O. Box 912, Beijing 100083, China}
\affiliation{CAS Center for Excellence in Topological Quantum Computation, University of Chinese Academy of Sciences, Beijing 100190, China}
\affiliation{Beijing Academy of Quantum Information Sciences, Beijing 100193, China}

\author{Justin C. W. Song}
\email{justinsong@ntu.edu.sg}
\affiliation{Division of Physics and Applied Physics, Nanyang Technological University, Singapore 637371}
\affiliation{Institute of High Performance Computing, Agency for Science, Technology, \& Research, Singapore 138632}

\begin{abstract}
The nonlinear shift current, also known as the bulk photovoltaic current generated by linearly polarized light, has long been known to be absent in crystals with inversion symmetry.
Here we argue that a non-zero shift current in centrosymmetric crystals can be activated by a photon-drag effect.
Photon-drag shift current proceeds from a `shift current dipole' (a geometric quantity characterizing interband transitions) and manifests a purely {\it transverse} response in centrosymmetric crystals.
This transverse nature proceeds directly from the shift-vector's pseudovector nature under mirror operation and underscores its intrinsic geometric origin.
Photon-drag shift current can greatly enhanced by coupling to polaritons and provides a new and sensitive tool to interrogate the subtle interband coherences of materials with inversion symmetry previously thought to be inaccessible via photocurrent probes.
\end{abstract}

\maketitle

The bulk photovoltaic effect (BPVE) produces a photocurrent in a single-phase homogeneous material~\cite{Chynoweth1956,Chen1969,Glass1974,Belinicher1980,Sturman1992,Fridkin2001} that persists even in the absence of conventional $p$-$n$ junctions.
This renders an entire bulk material active in photocurrent generation.
A prominent example of BPVE is the nonlinear shift current~\cite{Baltz1981,Young2012,Tan2016,Tan2016npj,Rangel2017,Nakamura2017,Cook2017,Yang2018,Burger2019} wherein geometric phases sustained by electronic states~\cite{Morimoto2016,Wang2019} enable a photo-induced current in the bulk.
While such geometric phases can be found in a large variety of materials, since photocurrent is a vector, broken (intrinsic) symmetries are required in fixing the direction of shift current in a uniform bulk. As a result, shift currents are typically thought to vanish in centrosymmetric materials~\cite{Belinicher1980,Sturman1992,Fridkin2001,Tan2016npj}, even in those possessing non-trivial geometric phases.

\begin{figure}[t!] 
\includegraphics[width=0.95\columnwidth]{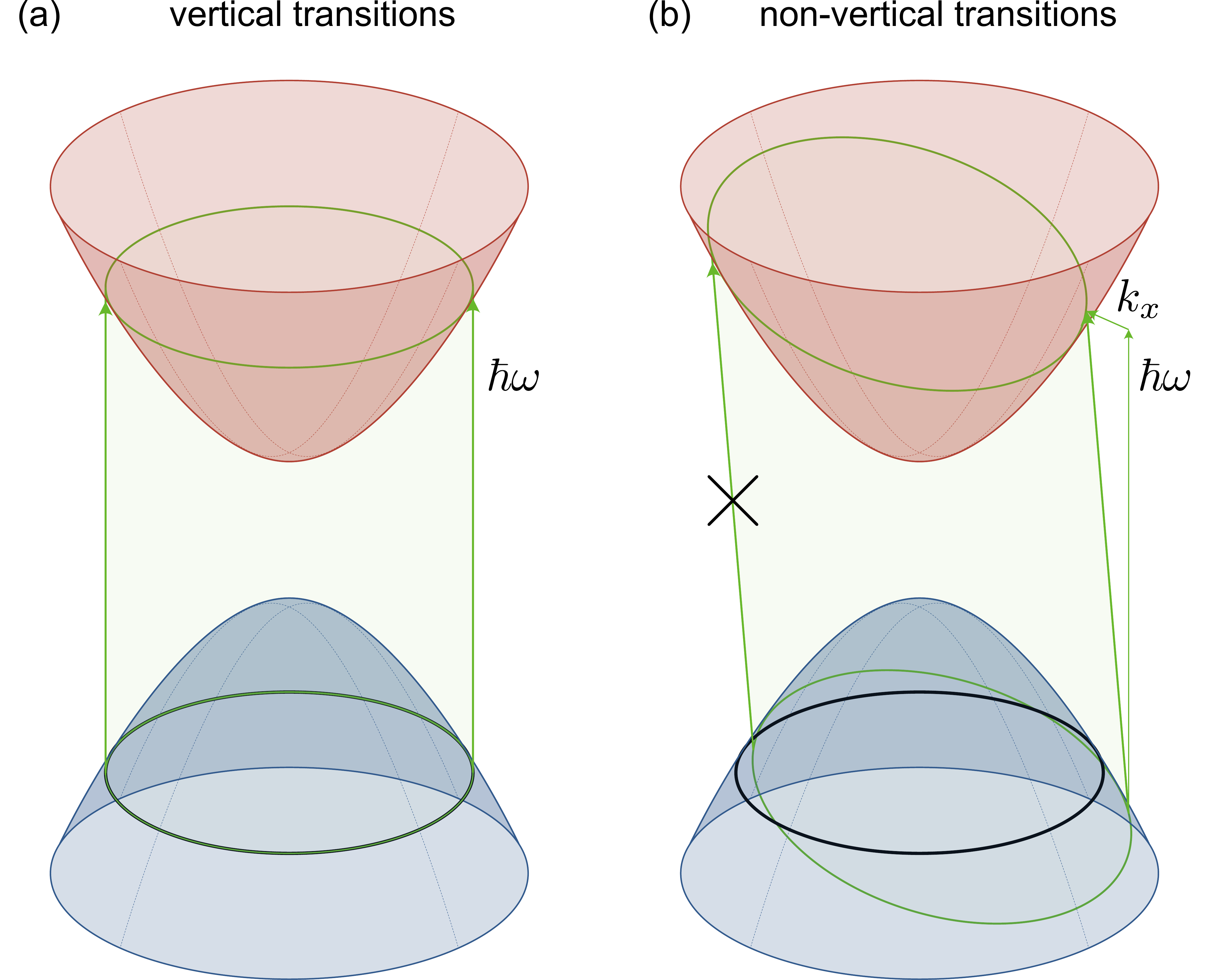}
\caption{Schematic comparison between (a) vertical transitions and (b) non-vertical transitions (e.g., from photon-drag effects) between two bands.
Solid black contours denotes the Fermi surface (FS) crossing the valence band (blue pockets). Green contours are energy-momentum-conserving contours (ECs) that satisfies (a) vertical transitions $\omega_{ {\rm c}, \vec p} - \omega_{ {\rm v}, \vec p }  - \omega = 0$ or (b) non-vertical transitions $\omega_{ {\rm c}, \vec p + k_x/2} - \omega_{ {\rm v}, \vec p - k_x/2 }  - \omega = 0$. For non-vertical transitions, only part of the EC that below the FS are optically-allowed (right green arrow).}
\label{fig-tilted}
\end{figure}

Here we show that non-zero shift currents can be revived in centrosymmetric crystals.
In particular, we find that non-vertical transitions [Fig.~\ref{fig-tilted}(b)], readily enabled by photon/polariton-drag processes, produce finite shift currents even when crystal inversion symmetry remains unbroken. 
While requiring a finite momentum transfer, such photon-drag shift currents are intrinsic with a magnitude controlled by a `shift-current dipole' that captures the interband geometry present in a material; this closely parallels the Berry curvature dipole~\cite{Sodemann2015} describing intraband geometry.

Surprisingly, photon-drag shift currents are transverse to the momentum transfer in isotropic crystals with a longitudinal incident polarization.
As we explain below, this transverse nature arises from an intrinsic helical winding of electronic states found in many (centrosymmetric) systems (e.g., HgTe quantum wells, monolayer WTe$_2$, graphene) and vividly displays its geometric origin. 
This intrinsic behavior sharply contrasts with conventional photon-drag in isotropic crystals that is parallel/anti-parallel to the momentum transfer for a longitudinal polarization without angular momentum~\cite{Ribakovs1977,Grinberg1988,Ivchenko1978,Ivchenko2002,Ganichev2001,Hatano2009,Karch2010,Proscia2016,Akbari2017,Akbari2018,Strait2019}. 

We expect that photon-drag shift current can be found in a wide variety of centrosymmetric materials and can serve as a sensitive diagnostic of their interband geometry using readily available photocurrent spectroscopy -- previously thought impossible~\cite{Belinicher1980,Sturman1992,Fridkin2001,Tan2016npj}. This opens up a vast set of centrosymmetric materials to realize geometrical photocurrents. 

{\it Shift current and photon-drag ---}
First analyzed by von Baltz and Kraut~\cite{Baltz1981}, the shift current (density) arises from real-space displacements of an electron accrued during a photo-induced transition from an initial to final state:
\be
\vec j^{\rm s} = e \sum_{i \to f} W_{i \to f} \vec r_{i \to f},
\label{eq:general}
\ee
where $i$ and $f$ denote initial and final electronic states in momentum space, $W_{i \to f}$ is the photo-induced transition (absorption) rate, and $\vec r_{i \to f} $ is the real-space displacement acquired during the transition.

For a vertical optical transition between a valence (v) and conduction (c) band, an electron's initial and final state has the same momentum, $\vec p$ [see Fig.~\ref{fig-tilted}(a)]. As a result, $W_{i \to f} = W(\vec p, {\rm v} \to {\rm c})$, with $\vec r_{i\to f}$ described by the shift vector~\cite{Baltz1981,Sipe2000,Sinitsyn2006,Shi2019}
\be
\vec r^{(0)} (\vec p) = {\vec A}_{\rm c} (\vec p) - {\vec A}_{\rm v} (\vec p) - \nabla_{\vec p} \arg [ \nu^{(0)} (\vec p) ] ,
\label{eq:originalshiftvector}
\ee
where ${\vec A}_{\rm c,v} ( \vec p ) = \la u_{\rm c,v} (\vec p) | i \nabla_{\vec p} u_{\rm c,v} (\vec p) \ra $ is the Berry connection for c,v bands, and $ \nu^{(0)} (\vec p) =  \la u_{\rm c} (\vec p) | \hat{\nu} | u_{\rm v} (\vec p) \ra $ is a velocity matrix element. The form of $\hat{\nu}$ is determined by the polarization of light and the electronic Hamiltonian.  

For crystals with inversion symmetry, the shift vector $\vec r^{(0)} (\vec p) = - \vec r^{(0)} ( - \vec p)$ is odd in momentum space (see e.g., below and in the Supplementary Information, {\bf SI}). In contrast, the transition rate $W(\vec p, {\rm v } \to {\rm c} )$ is even under inversion. As a result the shift current density Eq.~(\ref{eq:general}) vanishes in centrosymmetric crystals~\cite{Belinicher1980,Sturman1992,Fridkin2001,Tan2016npj}.

As we now argue, this constraint can be circumvented even in centrosymmetric crystals by considering non-vertical transitions, shown in Fig.~\ref{fig-tilted}(b).
Such non-vertical transitions readily manifest from photon-drag (or polariton-drag, see below) which include momentum transfer from photons to electrons: the initial and final states read as $ | u_{\rm v} (\vec p-\vec k/2 ) \ra $ and $ | u_{\rm c} (\vec p + \vec k/2 ) \ra $ with $\hbar \vec k$ the momentum transferred from the photon to the electron~\cite{Shalygin2016}. Using Eq.~(\ref{eq:general}) and Fermi's golden rule, we obtain a photon-drag shift current density as
\be
\vec j^{\rm s} (\vec k) = C \int_{\vec p} 
 \rho (\vec p, \vec k) \vec R^{} (\vec p, \vec k), \,\, 
\vec R^{} (\vec p, \vec k) \equiv | \nu (\vec p,\vec k) |^2  \vec r (\vec p, \vec k) ,
\label{eq:jk}
\ee
where $C = e (\pi/2) (e E / \hbar \omega)^2 $
contains the electric field strength $E$ and light frequency $\omega$, the $d$-dimensional integral is written as $ \int_{\vec p} \equiv \int {\rm d}^d p_i / (2 \pi)^d$, and $\rho (\vec p, \vec k) = [ f (\epsilon_{ {\rm v}, \vec p - \vec k/2 }) - f (\epsilon_{ {\rm c}, \vec p + \vec k/2 }) ] \delta ( \omega_{ {\rm c}, \vec p + \vec k/2 } - \omega_{ {\rm v}, \vec p-\vec k/2 }  - \omega ) $ defines a tilted, optically-allowed {\it energy-momentum-conserving contour} (EC) in momentum space [see Fig.~\ref{fig-tilted}(b)].
Without loss of generality, we will focus on light polarized along the $x$-axis so that the velocity matrix element is $ \nu (\vec p, \vec k) =  \la u_{\rm c} (\vec p + \vec k/2 ) | \hat{\nu}_x | u_{\rm v} (\vec p-\vec k/2 ) \ra $ where $\hat{\nu}_x = \p \hat{\cal H} (\vec p) / \hbar \p p_x$.  Crucially, real-space displacements $\vec r_{i\to f}$ for non-vertical transitions in Fig.~\ref{fig-tilted}(b) are
\be
\vec r (\vec p, \vec k) = {\vec A}_{\rm c} (\vec p + \vec k/2 ) - {\vec A}_{\rm v} (\vec p-\vec k/2) - \nabla_{\vec p} \arg [ \nu (\vec p, \vec k) ] .
\label{eq:shift-vector}
\ee
When $\vec k = \vec 0$, $\vec j^{\rm s} (\vec 0) $ in Eq.~(\ref{eq:jk}) reduces to the conventional shift current for vertical transitions without photon-drag. Indeed, $\vec r (\vec p, \vec k = \vec 0) = \vec r^{(0)} (\vec p) $ in Eq.~(\ref{eq:originalshiftvector}).

On a fundamental level, we note that the shift in Eq.~(\ref{eq:shift-vector}) [as well as Eq.~(\ref{eq:originalshiftvector})] is in fact a geometrical quantity that captures an inter-band geometry between the conduction and valence bands.
Note that the interband transitions in Fig.~\ref{fig-tilted} accumulates a gauge invariant (interband) phase (technically, the phase of a Wilson-loop associated with the transition). The shift in Eq.~(\ref{eq:shift-vector}) is the gradient of such an interband phase associated with the transition~\cite{Shi2019}, see {\bf SI}. This mirrors how local (intraband) Berry curvature at $\vec p$ captures the Berry phase accrued over an infinitesimally small loop around $\vec p$~\cite{Fukui2005}. 

Macroscopically, the photon-drag shift current can be written as $ j_\beta^{\rm s} = E_\alpha^2 \sigma_{\alpha \alpha \beta}^{\rm s} (\vec k) $. For centrosymmetric crystals, the inversion operation leads to $ j_\beta^{\rm s} \to - j_\beta^{\rm s} $, $E_{\alpha}^2 \to E_{\alpha}^2$, and $ \sigma_{\alpha \alpha \beta}^{\rm s} (\vec k) \to \sigma_{\alpha \alpha \beta}^{\rm s} ( - \vec k)$. As long as $\sigma_{\alpha \alpha \beta}^{\rm s} ( - \vec k) \neq \sigma_{\alpha \alpha \beta}^{\rm s} (\vec k) $, one can expect the emergence of $\sigma_{\alpha \alpha \beta}^{\rm s} (\vec k)$ as well as $j_\beta^{\rm s}$ [as captured in Eq.~(\ref{eq:jk})].
A seemingly natural expectation is that the such a photon-drag shift current should be parallel to $\vec k$, because (1) from the point of view of shift current, this is the direction where the overall centrosymmetry (photon + crystal) is broken; (2) from the perspective of photon-drag effect, this is the direction where the momentum transfer happens.
However, as we show, a non-zero shift current induced by photon-drag can be {\it transverse} to $\vec k$ due to its geometric property.

{\it Shift current dipole ---}
In order to relate the photon-drag shift current $\vec j^{\rm s} (\vec k) $ to the intrinsic properties of a centrosymmetric crystal, we expand $j_\alpha^{\rm s} (\vec k)$ at small $\vec k$, representing it as a product of $\vec k$ with a `shift current dipole', $\vec D$, via 
\be
j_\beta^{\rm s} (\vec k) = k_\alpha D_{\alpha \beta} + O (k^2),
\label{eq:jk2}
\ee
where $\alpha,\beta = x,y,z$, and repeated indices are implicitly summed over.
The shift current dipole is 
\begin{equation}
D_{\alpha \beta} =C \int_{\vec p} [ d_\alpha^\rho (\vec p) 
R_\beta^{(0)} (\vec p)
+ d_{\alpha \beta}^R (\vec p) \rho^{(0)} (\vec p) ] ,
\label{eq:shiftcurrentdipole}
\end{equation}
where $R_\beta^{(0)} (\vec p)$ and $\rho^{(0)} (\vec p) $ are obtained at $\vec k = \vec 0$ similar to $\vec r^{(0)} (\vec p)$, $d_{\alpha \beta}^R (\vec p) = [ \p R_\beta (\vec p, \vec k) /  \p k_\alpha ]_{\vec k = \vec 0} $,
and
\begin{align}
d_\alpha^\rho (\vec p) = \frac{\p \rho (\vec p, \vec k)}{\p k_\alpha} \Big|_{\vec k = \vec 0}  
=  - \frac{1}{2} \frac{\p (  f_{{\rm v}, \vec p } + f_{{\rm c}, \vec p } ) }{ \p  p_\alpha }   \,  \delta ( \omega_{\vec p}^{\rm cv} - \omega ) ,
\label{eq:d-alpha-rho}
\end{align}
with $ f_{ {\rm c (v)}, \vec p } = [ 1 + e^{ ( \epsilon_{ {\rm c (v)}, \vec p } - \mu ) / k_B T } ]^{-1} $ the Fermi function and $\omega_{\vec p}^{\rm cv} = \omega_{ {\rm c}, \vec p} - \omega_{ {\rm v}, \vec p}  $ (see {\bf SI} for a detailed derivation).

We note that $R_\beta^{(0)} (\vec p) = | \nu^{(0)} (\vec p) |^2  r^{(0)}_\beta (\vec p)$ is odd with respect to $\vec p \to - \vec p$ [as expected from $\vec r^{(0)} (\vec p)$ in a centrosymmetric crystal, Eq.~(\ref{eq:shift-vector})]. Similarly $d_\alpha^\rho (\vec p)$ is controlled by the group velocity along $\alpha$ [Eq.(\ref{eq:d-alpha-rho})] and is also odd. As a result, the first term of Eq.~(\ref{eq:shiftcurrentdipole}) yields a non-zero contribution to the shift current dipole $ D_{\alpha \beta}$.
On the other hand, since $\rho^{(0)} (\vec p)$ is even with $\vec p$, the second term of Eq.~(\ref{eq:shiftcurrentdipole}) measures the evenness of $d_{\alpha \beta}^R (\vec p)$.
We note, parenthetically, that the second term in Eq.~(\ref{eq:shiftcurrentdipole}) vanishes when a centrosymmetric crystal possesses an additional particle-hole symmetry (PHS) (common in low energy two-band systems). This is because PHS gives rise to an even $R_\beta (\vec p, \vec k) = R_\beta (\vec p, - \vec k)$ (see {\bf SI}) and a vanishing $d_{\alpha \beta}^R (\vec p) = 0$. As a result, in what follows, we will concentrate on the first term which typically dominates.

{\it Symmetry and transverse nature of shift current dipole ---}
From Eqs.~(\ref{eq:shift-vector}) and (\ref{eq:shiftcurrentdipole}), the shift current dipole depends on both the intrinsic properties of the crystal and the light polarization. For a generic centrosymmetric crystal without any additional crystalline symmetry, both $D_{x x}$ and $D_{x y}$ can be non-zero. However, as we now show, crystalline symmetry can severely constrain the form of the shift current dipole $D_{\alpha \beta}$.

To see this, we analyze the effect of time reversal symmetry (TRS) and mirror symmetry (MS) in a 2D centrosymmetric crystal. 
We first focus on incident light with linear polarization [captured in $\hat{\nu}_x$] and its wave vector $\vec k = k_x \hat{\vec x}$ both being parallel to the mirror plane (fixed along the $x$-axis); in this case, the light electric field does not break overall MS. For centrosymmetric crystals with TRS and MS (${\cal M}_y: y \to -y$), the spin-resolved shift vector $[\vec r^{(0)} (\vec p)]^\sigma$ satisfy symmetry constrained relations; here $\sigma = \uparrow, \downarrow$. These symmetry constraints can be readily obtained by directly analyzing how Eq.~(\ref{eq:originalshiftvector}) (or equivalently, the Wilson-loop associated with the interband transitions) transforms under time-reversal and mirror operations, see details in {\bf SI}.
First, we find
\be
[\vec r^{(0)}(\vec p)]^\uparrow = [\vec r^{(0)}(-\vec p)]^\downarrow = - [\vec r^{(0)}(\vec p)]^\downarrow,
\label{eq:r-TRIS}
\ee
where the first equality comes from TRS while the second equality arises from inversion symmetry (IS). In the presence of MS, we have
\be
[r^{(0)}_{x} (\vec p)]^ \sigma = - [r^{(0)}_{x} ({\cal M}_y \vec p)]^ \sigma  ,
~
[r^{(0)}_{y}(\vec p)]^ \sigma  =  [r^{(0)}_{y} ({\cal M}_y \vec p)]^ \sigma  .
\label{eq:pseudo-vector}
\ee
In obtaining Eq.~(\ref{eq:pseudo-vector}) we have repeatedly applied MS, TRS, and IS. Interestingly, Eq.~(\ref{eq:pseudo-vector}) means that $[\vec r^{(0)} (\vec p)]^\sigma$  behaves as a pseudo-vector with respect to the mirror plane. 
Noting that $|[\nu^{(0)} (\vec p)]^{\sigma}|^2$ is invariant under the same TR and Mirror operations as above, we find that Eq.~(\ref{eq:r-TRIS}) and (\ref{eq:pseudo-vector}) also hold when we replace $\vec r^{(0)} (\vec p) \to \vec R^{(0)} (\vec p)$.
As a result, $[ \vec R^{(0)} (\vec p) ]^\sigma $ in Eq.~(\ref{eq:shiftcurrentdipole}) also acts as a pseudo-vector.

Since $d_x^\rho (\vec p)  = d_x^\rho ( {\cal M}_y \vec p)$ in Eq.~(\ref{eq:d-alpha-rho}) is even about the mirror plane, the pseudo-vector nature of
$[ \vec R^{(0)} (\vec p) ]^\sigma $ [see Eq.~(\ref{eq:pseudo-vector})] enforces a {\it vanishing} $D_{x x}^\sigma = 0$, but allows a finite $D_{x y}^\sigma \neq 0$. As a result, the shift current dipole is purely {\it transverse}. 
Interestingly, even when TRS and MS are individually broken, as long as their composite symmetry operation ${\cal O} = {\cal M}_y {\cal T} $ is present, Eq.~(\ref{eq:pseudo-vector}) persists and the pseudo-vector nature of
$[ \vec R^{(0)} (\vec p) ]^\sigma $ is preserved (see {\bf SI}) yielding a purely transverse $\vec D^\sigma$.

We note that when the electronic system possess an effective $U(1)$ continuous rotational symmetry, then any in-plane axis also acts as a mirror axis. In such a circumstance, $D_{\alpha \beta}^\sigma$ is similarly purely transverse for linear polarizations applied along any in-plane axis.
However, so long as linear polarization is not directed along a mirror plane (overall mirror symmetry is broken), all components of the shift-current dipole $D_{\alpha \beta}^\sigma$ are generically allowed.

\begin{figure}[t!] 
\includegraphics[width=1\columnwidth]{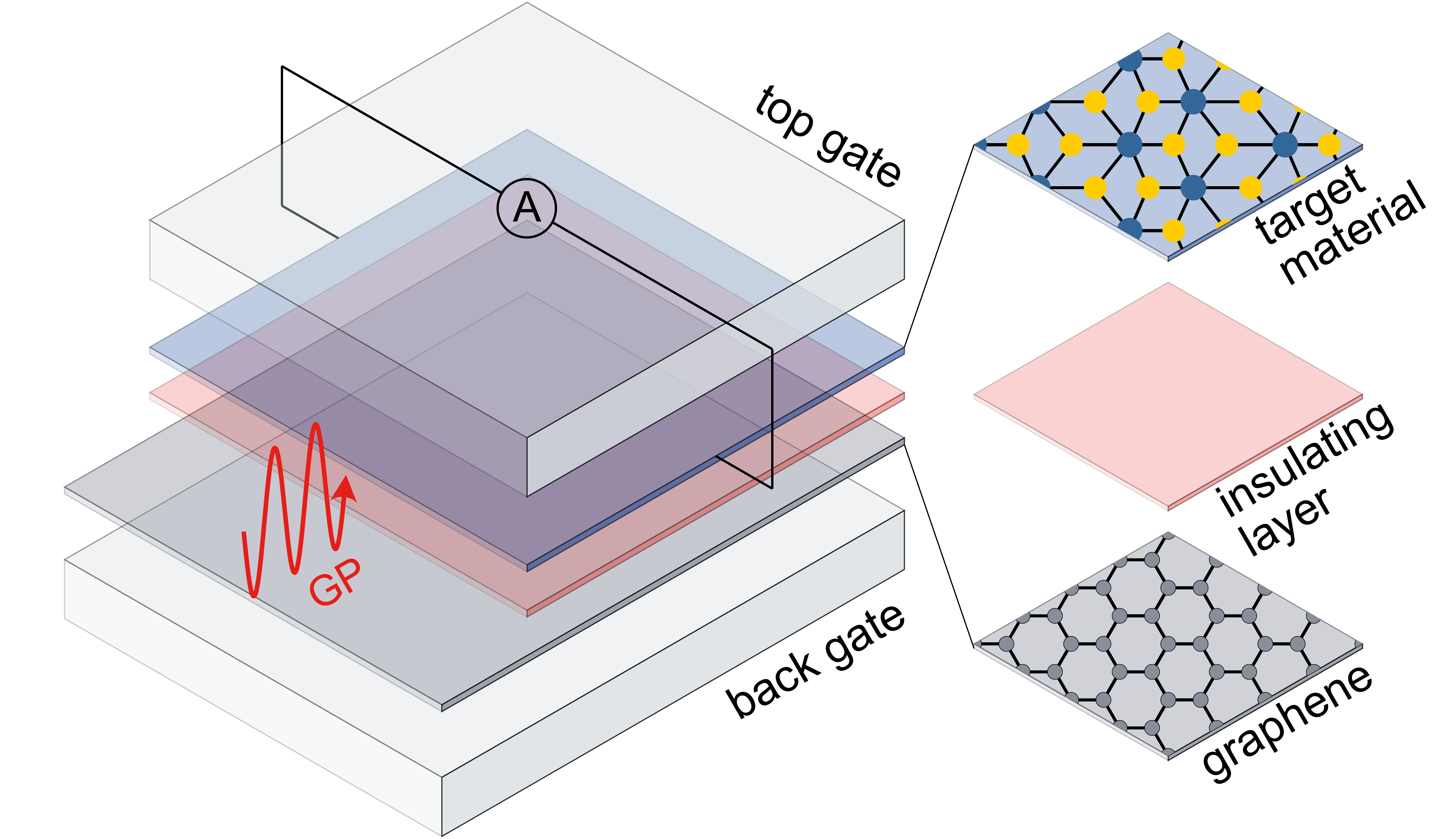}
\caption{Schematic of graphene plasmon enhancing the photon-drag effect. A centrosymmetric 2D or thin-film target material (blue layer), is stacked on top of a graphene monolayer (thin gray layer), with an insulating layer (purple layer) in between. A propagating graphene plasmon (GP, red curve) can induce non-vertical optical transitions and generates a shift current in the adjacent target material, the later of which can be perpendicular to the GP propagation direction, i.e., a transverse photo-drag effect. Fermi surface of the target crystal is tunable with top and bottom gates (thick gray layers).}
\label{fig-enhance}
\end{figure}

{\it Polariton enhanced photon-drag ---}
In most cases, the wavelength (wavevector $\vec k$) of light is much larger (smaller) than typical electron wavelengths (wavevectors).
As a result, the photon-drag shift current [see Eq.~(\ref{eq:jk2})] expected can be small. However, as we show below, sizeable $j_\beta^{\rm s} (\vec k)$ can be achieved when coupling with polaritons~\cite{Basov2016,Ni2018} which have a much slower speed than photons massively amplifying the wavevector $\vec k$ at the same frequency~\cite{Kurman2018}.
Graphene plasmons (GP) are exceptionally tailored to achieve this task because (1) its wavelength can be masssively compressed to $50$ to $100$ nm within a large frequency window~\cite{Basov2016} (compression factors can be as large as 300);
(2) GP can possess a large quality factor (as large as 130)~\cite{Ni2018} and can propagate through and cover a large sample; (3) GP generates a strong AC electric field that can extend out to its surrounding environment (this extent is of order the GP plasmon wavelength).

As such, a layered van der Waals stacked structure (see Fig.~\ref{fig-enhance}), can be readily employed to plasmonically enhance the photon-drag shift current in a target 2D material.
By stacking a 2D or thin-film target crystal on top of a graphene layer with a thin insulating layer (e.g., hexagonal Boron Nitride that can be as thin as several nm) in between, then exciting a propagating GP in the graphene, the longitudinal AC electric field generated by the GP~\cite{Chen2012,Fei2012} (whose linear polarization aligns with its large wave vector $\vec k$) can trigger non-vertical transitions~\cite{Pratama2019} in the target layer.

{\it Spin and charge transverse photon-drag shift current ---} We now turn to exemplify the photon-drag shift current in a minimal low-energy model of a 2D centrosymmetric crystal, the Bernevig-Hughes-Zhang (BHZ) model~\cite{Bernevig2006}
\be
{\cal H}_0 = m_{\vec p} s_0 \tau_z + v_x p_x  s_z \tau_x - v_y p_y s_0 \tau_y, 
\label{eq:bhz}
\ee
describing two spin-degenerate bands with TRS, where $s_{x,y,z}$ and $\tau_{x,y,z}$ denote spin and orbital degrees of freedom, respectively; $s_0$ and $\tau_0$ are $2\times2$ identity matrices, $m_{\vec p} = - m_0 + c_x p_x^2 + c_y p_y^2$. The BHZ model can characterize the low-energy electronic and optical behavior of a wide variety of systems. For example, when $c_x = c_y, v_x = v_y$ it captures an isotropic and centrosymmetric electronic system (e.g., HgTe quantum wells~\cite{Bernevig2006}) and when $c_x \neq c_y$ or $v_x \neq v_y$ it describes a centrosymmetric system with a single mirror plane~\cite{ShiWTe2}.
To clearly exhibit the pseudovector nature of $[ \vec R^{(0)} (\vec p) ]^\sigma $, we concentrate on the latter case with a mirror plane along $x$-axis.

\begin{figure}[t!] 
\includegraphics[width=1\columnwidth]{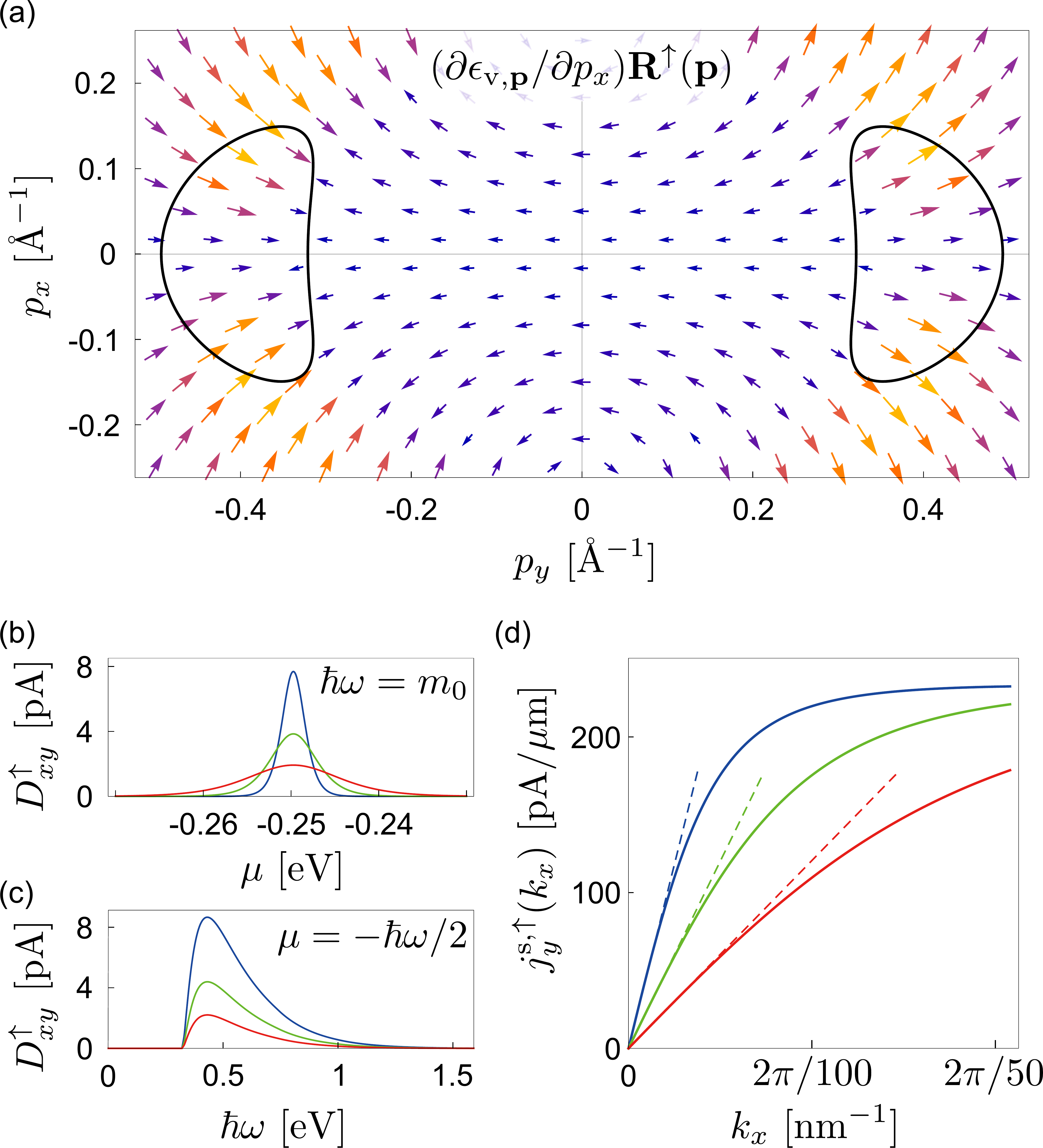}
\caption{(a) Calculated pseudo-vector field $( \p \epsilon_{{\rm v}, \vec p} / \p p_x) \vec R^{\uparrow} (\vec p) $ for the BHZ model at $T=0$, with the black contours denote the integrated region $\delta (\omega_{\vec p}^{\rm cv} - \omega ) 
\delta ( \epsilon_{ {\rm v}, \vec p } - \mu )$ contributing to the shift current dipole [see Eq.~(\ref{eq:dipoleT0})]. The relative sizes and colors denote the relative magnitude for each of the vector.
Shift current dipole (b) as a function of chemical potential $\mu$ at a fixed plasmon energy $\hbar \omega = 500~{\rm meV}$, and (c) as a function of plasmon energy $\hbar \omega$ with a fixed relation $\mu = - \hbar \omega / 2$ at different temperatures ($T= 10, 20, 40 ~{\rm K}$ for blue, green, red curves).
(d) Shift current calculated directly from Eq.~(\ref{eq:jk}) (solid curves) versus linear approximations from shift current dipole (dashed lines).
Parameter used for the two-band model: $m_0 = 0.5~{\rm eV}$, $c_x = 6~{\rm eV\, \AA^{-2}}$, $c_y = 3~{\rm eV\, \AA^{-2}}$, $v_x = 1.4~{\rm eV\, \AA^{-1}}$, $v_y = 0.4~{\rm eV\, \AA^{-1}}$, these correspond to values found in monolayer WTe$_2$~\cite{ShiWTe2}. Here we used a GP electric field with $E = 1000~{\rm V \,cm^{-1}}$~\cite{Pratama2019}.}
\label{fig-bhz}
\end{figure}

We first focus on the spin-up branch of Eq.~(\ref{eq:bhz}) and calculate its shift current dipole, see parameter values for Eq.~(\ref{eq:bhz}) in caption. Other parameters values can also be used with no qualitative change to our results. We note that $ {\cal H}_0^{\uparrow} $ possesses PHS yielding $d_{\alpha \beta}^{R, \uparrow} (\vec p) = 0$. At zero temperature $T=0$ and assuming the chemical potential crosses the valence band, the shift current dipole for the spin-up branch can be written as
\begin{align}
D_{\alpha \beta}^{\uparrow} =  - \frac{C}{2} \int_{\vec p} & \frac{\p \epsilon_{ {\rm v}, \vec p } }{ \p  p_\alpha } 
R_\beta^{\uparrow} (\vec p) ~ 
\delta ( \omega_{\vec p}^{\rm cv} - \omega ) 
\delta ( \epsilon_{ {\rm v}, \vec p } - \mu ), 
\label{eq:dipoleT0}
\end{align}
where we have dropped mention of $\vec k = \vec 0$ in $\vec R$ for brevity.
We plot $ ( \p \epsilon_{ {\rm v}, \vec p } / \p  p_x ) \vec R^{\uparrow} (\vec p) $ in Fig.~\ref{fig-bhz}(a), which shows that $\vec R^{\uparrow} (\vec p)$ as well as $ ( \p \epsilon_{ {\rm v}, \vec p } / \p  p_x ) \vec R^{\uparrow} (\vec p) $ are indeed pseudo-vector fields with respect to the mirror plane along $x$-axis. By integrating over the EC denoted by $\delta ( \omega_{\vec p}^{\rm cv} - \omega ) 
\delta ( \epsilon_{ {\rm v}, \vec p } - \mu )$ [black contours in Fig.~\ref{fig-bhz}(a)], we obtain a vanishing $D_{x x}^{\uparrow}$ and a non-zero $D_{x y}^{\uparrow}$. This gives a purely transverse $D_{\alpha \beta}^{\uparrow} $ as expected from the above symmetry analysis.

We now turn to the finite temperature behavior of $D_{x y}^{\uparrow}$. In Fig.~\ref{fig-bhz}(b), we fix the polariton frequency (GP frequency) $\hbar \omega_0$ and plot $D_{x y}^{\uparrow}$ in Eq.~(\ref{eq:shiftcurrentdipole}) for $\mathcal{H}_0$ in Eq.~(\ref{eq:bhz}) for various the chemical potentials and temperature values. Fig.~\ref{fig-bhz}(b) displays a peaked $D_{x y}^{\uparrow}$ [with a width over a sizeable energy window: 10 to 20 meV] representing pronounced $D_{x y}^{\uparrow}$ when the Fermi surface crosses the EC; similarly, the width increases as temperature increases. Fig.~\ref{fig-bhz}(b) also shows that peak $D_{x y}^{\uparrow}$ appears at $\mu = - \hbar \omega / 2$, i.e., when FS and EC overlap greatest with each other. Guided by this latter observation, we fix $\mu = - \hbar \omega / 2 $ and compute $D_{x y}^{\uparrow}$ as a function of $\hbar \omega$ in Fig.~\ref{fig-bhz}(c). Strikingly, $D_{x y}^{\uparrow}$ manifests over a wide window of photon/polariton energies.

In Fig.~\ref{fig-bhz}(d), we compare the up-spin photon-drag shift current $j_y^{\rm s, \uparrow} (k_x)$ calculated from a direct integration from Eq.~(\ref{eq:jk}) and that from linear approximation in Eq.~(\ref{eq:jk2}).
For small $k_x$, $j_y^{\rm s, \uparrow} (k_x)$ grows linearly with $k_x$ as expected from Eq.~(\ref{eq:jk2}) [dashed and solid lines coincide]. At low temperatures, the transverse shift current begins to saturate large $k_x$ ($\lambda_p < 100$ nm); at higher temperatures, the linear region becomes wider due to a temperature smeared out Fermi surface. Saturation arises when EC maximal tilts $|\vec k| \gtrsim |\vec p|$.

We note that due to IS and TRS in Eq.~(\ref{eq:bhz}), $\vec R^{\uparrow} (\vec p) = - \vec R^{\downarrow} ( \vec p)$. As a result, $\vec j^{\rm s, \uparrow} (k_x) = -\vec j^{\rm s, \downarrow} (k_x)$ leading to spin photon-drag shift currents that propagate in opposite directions for spin $\uparrow, \downarrow$ but a vanishing charge photon-drag shift current. The former spin currents propagate transverse to $\vec k$. 

A finite {\it charge} photon-drag shift current, however, can be readily revealed when TRS is broken.
One way to achieve this is via a Zeeman effect that splits the spin degeneracy:
${\cal H} = {\cal H}_0 + {\cal H}_B$ with
\be
{\cal H}_B = \Delta s_z \tau_0 .
\ee
This additional Zeeman term can be induced in two ways. Namely, directly applying a perpendicular magnetic field which splits electrons with opposite spins, as well as
stacking another layer of magnetic material such as CrI$_3$ on top of the target crystal~\cite{Zhao2020}. In the latter, exchange interactions can similarly break the TRS of the target material ${\cal H}_0$~\cite{Liu2008,Chang2013}. Broken TRS imbalances $\vec j^{\rm s, \uparrow} (k_x)$ and $\vec j^{\rm s, \downarrow} (k_x)$ leading to a finite charge photon-drag shift current. 

For large TRS breaking so that spin-up and spin-down branches are separated by more than the $D_{x y}^{\uparrow}$  width shown in Fig.~\ref{fig-bhz}(b), the optical responses are then determined by electrons from a single spin branch only, e.g., $\vec R(\vec p) = \vec R^\uparrow (\vec p)$. Importantly, we emphasize that ${\cal H}_0 + {\cal H}_B$ still preserves the composite symmetry ${\cal O} = {\cal M}_y {\cal T} $. As a result, $\vec R (\vec p)$ remains a pseudovector guaranteeing a purely transverse $D_{\alpha \beta}$ and transverse $\vec j^s$.

Photon-drag shift current is a geometric effect that proceeds directly from the subtle wavefunction coherences between conduction and valence bands. Arising even in centrosymmetric crystals, it can be described by a shift-current dipole $D_{\alpha \beta}$ that quantifies the susceptibility of IS materials to {\it interband} (geometrical) effects; this parallels the Berry curvature dipole that captures the {\it intraband} nonlinear Hall effect in crystals with TRS. While we have focussed on the shift current (arising from linearly polarized light), another related geometric photocurrent can be induced by circularly polarized light -- the injection current. 
We anticipate that the injection current, that was previously thought to vanish in centrosymmetric systems, can also become similarly ``un-blocked'' via non-vertical transitions, and can naturally exhibit charge photocurrents even in the absence of additional TRS breaking. Lastly, we note that bulk 3D materials (or 3D material thin films) can also exhibit photon-drag shift current, such as magnetic Weyl semimetals with inversion symmetry, e.g., Mn$_3$Sn~\cite{nakatsuji2015}. In such a case, TRS is intrinsically broken without applying an external magnetic field or a proximal magnetic layer.

\begin{acknowledgments}
{\it Acknowledgments -- }
L-K.S. gratefully acknowledges helpful conversations with Inti Sodemann.
J.C.W.S. acknowledges support from the National Research Foundation (NRF), Singapore under its NRF fellowship programme award number NRF-NRFF2016-05, the Ministry of Education, Singapore under its MOE AcRF Tier 3 Award MOE2018-T3-1-002, and a Nanyang Technological University start-up grant (NTU-SUG). K.C. acknowledges support by the NSFC (Grants No.\,61674145), and the Chinese Academy of Sciences (Grants No.\,QYZDJ-SSW-SYS001).
\end{acknowledgments}

\clearpage
\newpage

\renewcommand{\theequation}{S-\arabic{equation}}
\renewcommand{\thefigure}{S-\arabic{figure}}
\renewcommand{\thetable}{S-\Roman{table}}
\makeatletter
\renewcommand\@biblabel[1]{S#1.}
\setcounter{equation}{0}
\setcounter{figure}{0}

\onecolumngrid

\subsection*{\large{Supplementary Information for\\
``Geometric photon-drag effect and nonlinear shift current in centrosymmetric crystals''}}

\bigskip

\subsection{Wilson loop formalism and symmetry properties for the shift vector}

This section describes the symmetry properties of the shift vector. In order to clearly display its symmetry properties, we directly show how the shift vector proceeds from phases accumulated during transition processes described in the main text. We begin by noting that the Berry connection $\vec A_{\rm c,v} (\vec p)$ in Eq.~(\ref{eq:shift-vector}) essentially encodes phases between different Bloch eigenstates $\la u_{\rm c,v} (\vec p) | u_{\rm c,v} (\vec p + \vec q) \ra = \exp [ - i \vec A_{\rm c,v} (\vec p) \cdot \vec q + {\cal O} (q^2) ] $, and can be expressed as $\vec A_{\rm c,v} (\vec p) = - {\rm lim}_{\vec q\to 0} \nabla_{\vec q} \arg [ \la u_{\rm c,v} (\vec p) | u_{\rm c,v} (\vec p + \vec q ) \ra ] |$.
Using this, we can rewrite $\vec r (\vec p, \vec k)$ in Eq.~(\ref{eq:shift-vector}) as the gradient of a phase
\be
\vec r (\vec p, \vec k)
=
\lim_{\vec q \to \vec 0}
\nabla_{\vec q} 
\arg [ {\cal W} ( \vec p, \vec q, \vec k) ] ,
\label{eq:shift-as-a-phase-gradient}
\ee
with the Wilson loop ${\cal W} ( \vec p, \vec q, \vec k)$ 
\begin{align}
{\cal W} ( \vec p, \vec q, \vec k) 
= \, &
\la u_{\rm v} (\vec p - \vec k/2) | u_{\rm v} (\vec p + \vec q - \vec k/2 ) \ra
\la u_{\rm v} (\vec p + \vec q - \vec k/2 ) | \hat{\boldsymbol \nu} | u_{\rm c} ( \vec p + \vec q + \vec k/2 ) \ra
\nn
& \cdot \la u_{\rm c} ( \vec p + \vec q + \vec k/2 )  | u_{\rm c} (\vec p + \vec k/2 ) \ra
\la u_{\rm c} (\vec p + \vec k/2 ) | u_{\rm v} (\vec p - \vec k/2 ) \ra  ,
\label{eq:whole-loop}
\end{align}
encoding the transition process $u_{\rm v} (\vec p - \vec k/2 ) \to u_{\rm v} (\vec p + \vec q - \vec k/2 ) \to \hat{\boldsymbol \nu} \to u_{\rm c} (\vec p + \vec q + \vec k / 2 ) \to u_{\rm c} (\vec p + \vec k /2 )$.
In obtaining Eq.~(\ref{eq:whole-loop}) we have added $ \la u_{\rm c } (\vec p + \vec k/2) | u_{\rm v} (\vec p - \vec k/2 ) \ra $ (last term) that is $\vec q$-independent to create a closed loop; its contribution to $\vec r (\vec p, \vec k)$ vanishes under the action of $\nabla_{\vec q}$ in Eq.~(\ref{eq:shift-as-a-phase-gradient}). We note that even without the last term, the first four terms of Eq.~(\ref{eq:whole-loop}) give a Wilson line that under the action of $\nabla_{\vec q}$ remains gauge invariant as all Bloch state vectors containing $\vec q$ always appear in pairs. The symmetry properties of $\vec r (\vec p, \vec k)$ are therefore determined by those of ${\cal W} ( \vec p, \vec q, \vec k) $.

The conventional shift vector, $\vec r^{(0)} (\vec p)$ (valid for vertical transitions), can also be directly obtained from Eq.~(\ref{eq:shift-as-a-phase-gradient}) as
\be
\vec r^{(0)} (\vec p) \equiv \vec r (\vec p, \vec 0)
= \lim_{\vec k, \vec q \to \vec 0}
\nabla_{\vec q} 
\arg [ {\cal W} ( \vec p, \vec q, \vec k) ]. 
\ee
This emphasizes the gauge invariant nature of $\vec r^{(0)} (\vec p)$ being a gradient of the phase obtained in the closed loop.

As discussed in the main text, we focus on centrosymmetric crystals with additional symmetries such as time reversal symmetry or mirror symmetry. For clarity, we assume that $\hat{\boldsymbol \nu} = \nu_x $ is along the high symmetry plane of the crystal (e.g., a mirror plan).

\bigskip

1)~ When a crystal has inversion symmetry, its full Hamiltonian in real-space $\cal H(\vec r)$ obeys the commutation relation $ [ {\cal H} (\vec r) , {\cal I}] = 0 $; here $ {\cal I} $ is the inversion operator. The Bloch Hamiltonian $ H (\vec k) = e^{-i \vec k \cdot \vec r} {\cal H} (\vec r) e^{ i \vec k \cdot \vec r}$ then satisfies
\be
{\cal I} H (\vec k) {\cal I}^{-1} = H (- \vec k) .
\label{eq:IS0}
\ee
The (Bloch) periodic eigenstates $| u_{n}^{\sigma} (\vec k) \ra$  ($\sigma = \, \uparrow, \downarrow$) of $H (\vec k)$ also inherit corresponding symmetry properties. To see this, we apply the inversion operator $\cal I$ on $\epsilon_{n, \vec k}^\sigma | u_{n}^{\sigma} (\vec k) $ and obtain
\be
{\cal I} [ \epsilon_{n, \vec k}^\sigma | u_{n}^{\sigma} (\vec k) ] =  {\cal I} [ H (\vec k) | u_{n}^{\sigma} (\vec k) ]
= {\cal I} H (\vec k) {\cal I}^{-1} | {\cal I}  u_{n}^{\sigma} (\vec k) \ra
= H (- \vec k) | {\cal I}  u_{n}^{\sigma} (\vec k) \ra ,
\label{eq:IS2}
\ee
where we used the symmetry relation Eq.~(\ref{eq:IS0}). On the other hand we have $ {\cal I} [ \epsilon_{n, \vec k}^\sigma | u_{n}^{\sigma} (\vec k) ]=
\epsilon_{n, \vec k}^\sigma | {\cal I}  u_{n}^{\sigma} (\vec k) \ra $ since $\epsilon_{n, \vec k}^\sigma$ is a real scalar. Together with Eq.~(\ref{eq:IS2}) we must have
\be
H (- \vec k) | {\cal I}  u_{n}^{\sigma} (\vec k) \ra = \epsilon_{n, \vec k}^\sigma | {\cal I}  u_{n}^{\sigma} (\vec k) \ra,
\ee
that leads to
\be
| {\cal I}  u_{n}^{\sigma} (\vec k) \ra = | u_{n}^{\sigma} ( - \vec k) \ra  ,
\quad
\epsilon_{n, \vec k}^\sigma = \epsilon_{n, - \vec k}^\sigma .
\label{eq:IS3}
\ee
Using Eq.~(\ref{eq:IS3}), we can readily see that
\be
\la u_{n_1}^{\sigma_1} (\vec v_1) | u_{n_2}^{\sigma_2} (\vec v_2) \ra
=
\la u_{n_1}^{\sigma_1} (\vec v_1) | {\cal I}^{-1} {\cal I}  | u_{n_2}^{\sigma_2} (\vec v_2) \ra
=
\la {\cal I} u_{n_1}^{\sigma_1} (\vec v_1) | {\cal I} u_{n_2}^{\sigma_2} (\vec v_2) \ra
= \la u_{n_1}^{\sigma_1} (- \vec v_1) | u_{n_2}^{\sigma_2} (- \vec v_2)  \ra ,
\label{eq:IS4}
\ee
and similarly,
\be
\la u_{n_1}^{\sigma_1} (\vec v_1) | \nu_x | u_{n_2}^{\sigma_2} (\vec v_2)  \ra 
= - \la u_{n_1}^{\sigma_1} (- \vec v_1) | \nu_x |u_{n_2}^{ \sigma_2} ( - \vec v_2)  \ra ,
\label{eq:IS5}
\ee
where we have used the fact that ${\cal I} \nu_x {\cal I}^{-1} = {\cal I} [ \p H(\vec k)/\p k_x ] {\cal I}^{-1} = \p H(- \vec k)/\p k_x = - \nu_x$, as readily obtained from Eq.~(\ref{eq:IS0}).

Eqs.~(\ref{eq:IS4}) and (\ref{eq:IS5}) guarantee that
\be
\arg[ {\cal W}^{\sigma} ( \vec p, \vec q, \vec k) ] = \arg [ {\cal W}^{\sigma} ( -\vec p, -\vec q, - \vec k) ] + \pi ,
\quad
[\vec r^{(0)}(\vec p)]^{\sigma} = - [\vec r^{(0)}(- \vec p)]^{\sigma} ,
\ee
where the argument function $\arg[z]$ here and below is defined within the interval $(-\pi, \pi]$.

\bigskip

2)~ When the crystal has both inversion symmetry $ {\cal I} $ and time reversal symmetry ${\cal T} =  -i \sigma_y K$, then we have 
\be
{\cal I} H (\vec k) {\cal I}^{-1} = H (- \vec k) ,
\quad
{\cal T} H (\vec k) {\cal T}^{-1} = H (- \vec k).
\label{eq:ITH}
\ee
Following similar analysis as above we have relations between (Bloch) periodic eigenstates
\be
| {\cal I}  u_{n}^{\sigma} (\vec k) \ra = | u_{n}^{\sigma} ( - \vec k) \ra ,
\quad
| {\cal T}  u_{n}^{\sigma} (\vec k) \ra = | u_{n}^{-\sigma} ( - \vec k) \ra^* .
\ee
as well as
\be
\la u_{n_1}^{\sigma_1} (\vec v_1) | u_{n_2}^{\sigma_2} (\vec v_2) \ra
= [ \la u_{n_1}^{- \sigma_1} (\vec v_1) | u_{n_2}^{- \sigma_2} (\vec v_2)  \ra ]^* ,
\quad
\la u_{n_1}^{\sigma_1} (\vec v_1) | \nu_x | u_{n_2}^{\sigma_2} (\vec v_2)  \ra 
= [ \la u_{n_1}^{- \sigma_1} (\vec v_1) | \nu_x |u_{n_2}^{ - \sigma_2} (\vec v_2)  \ra ]^*  ,
\label{eq:IS-TRS}
\ee
where we have consecutively carried out inversion and time reversal operations, used the fact that ${\cal I} \nu_x {\cal I}^{-1} = - \nu_x$ discussed above, and ${\cal T} \nu_x {\cal T}^{-1} = {\cal T} [ \p H(\vec k)/\p k_x ] {\cal T}^{-1} = \p H(- \vec k)/\p k_x = - \nu_x $ obtained from Eq.~(\ref{eq:ITH}).
Applying Eq.~(\ref{eq:IS-TRS}) in Eq.~(\ref{eq:whole-loop}), we see the spin-resolved Wilson loops obey
\be
\arg[ {\cal W}^{\sigma} ( \vec p, \vec q, \vec k) ] = \arg \big( [ {\cal W}^{- \sigma } ( \vec p, \vec q, \vec k) ]^* \big) = - \arg [ {\cal W}^{- \sigma} ( \vec p, \vec q, \vec k) ] ,
\ee
and the relation between the spin-resolved shift vectors
\be
[\vec r^{(0)}(\vec p)]^{\sigma} = - [\vec r^{(0)}(\vec p)]^{-\sigma} .
\ee

\bigskip

3)~ When the crystal has inversion symmetry $ {\cal I} $, time reversal symmetry ${\cal T} $, and mirror symmetry ${\cal M}_y $, we have
\be
| {\cal I}  u_{n}^{\sigma} (\vec k) \ra = | u_{n}^{\sigma} ( - \vec k) \ra ,
\quad
| {\cal T}  u_{n}^{\sigma} (\vec k) \ra = | u_{n}^{-\sigma} ( - \vec k) \ra^*,
\quad
| {\cal M}_y  u_{n}^{\sigma} (\vec k) \ra = | u_{n}^{-\sigma} ( {\cal M}_y \vec k) \ra,
\ee
and
\begin{align}
\la u_{n_1}^{\sigma_1} (\vec v_1) | u_{n_2}^{\sigma_2} (\vec v_2)  \ra 
& = [ \la u_{n_1}^{\sigma_1} ({\cal M}_y \vec v_1) | u_{n_2}^{\sigma_2} ({\cal M}_y \vec v_2) \ra ]^* ,
\nn
\la u_{n_1}^{\sigma_1} (\vec v_1) | \nu_x |u_{n_2}^{\sigma_2} (\vec v_2)  \ra 
& = [ \la u_{n_1}^{\sigma_1} ({\cal M}_y \vec v_1) | \nu_x| u_{n_2}^{\sigma_2} ({\cal M}_y \vec v_2) \ra ]^* ,
\label{eq:IS-TRS-My}
\end{align}
where we have consecutively carried out inversion, time reversal, and mirror (${\cal M}_y$) operations, used the fact that ${\cal I} \nu_x {\cal I}^{-1} = - \nu_x$ and ${\cal T} \nu_x {\cal T}^{-1} = - \nu_x $ discussed above, and ${\cal M}_y \nu_x {\cal M}_y^{-1} = \nu_x$.
In the same way as detailed above,  Eq.~(\ref{eq:IS-TRS-My}) applied on Eq.~(\ref{eq:whole-loop}) leads to
\be
\arg[ {\cal W}^{\sigma} ( \vec p, \vec q, \vec k) ] = - \arg [ {\cal W}^{\sigma} ( {\cal M}_y \vec p, {\cal M}_y \vec q, {\cal M}_y \vec k) ] ,
\ee
and
\be
[ r_{x}^{(0)} (\vec p)]^\sigma = - [ r_{x}^{(0)} ({\cal M}_y \vec p) ]^\sigma ,
\quad
[ r_{y}^{(0)} (\vec p) ]^\sigma = [ r_{y}^{(0)} ({\cal M}_y \vec p) ]^\sigma ,
\ee
which means that spin-resolved shift vectors are pseudo vectors with respect to the mirror plane.

\bigskip

4)~ When an external field breaks both TRS and MS, but preserves a composite symmetry ${\cal O} = {\cal M}_y {\cal T}$, then the pseudo-vector nature of $\vec r^\sigma (\vec p) $ persists because the Eq.~(\ref{eq:IS-TRS-My}) is still valid under this composite symmetry and inversion symmetry operations.

\bigskip

5)~ Apart from TRS and crystalline symmetries detailed above, another commonly seen non-spatial symmetry between conduction and valence bands is particle-hole symmetry (PHS).

For a generic two-band Hamiltonian (repeated indices are implicitly summed over)
\be
H (\vec k) = h_i (\vec k) \sigma_i ,
\quad
(i = x,y,z)
\ee
which has eigenenergies $ \epsilon_{\rm c,v} (\vec k) = \pm \sqrt{ h_i (\vec k) h_i (\vec k)} $ and eigenstates
\begin{equation}
| u_{\rm c} (\vec k) \ra =
\begin{bmatrix}
\cos ( \theta_{\vec k} / 2 ) e^{ - i \phi_{\vec k} / 2 }  \\
\sin  ( \theta_{\vec k} / 2 ) e^{ i \phi_{\vec k} / 2  }
\end{bmatrix}
,
\quad
| u_{\rm v} (\vec k) \ra =
\begin{bmatrix}
\sin ( \theta_{\vec k} / 2 ) e^{ - i \phi_{\vec k} / 2 } \\
- \cos  ( \theta_{\vec k} / 2 ) e^{ i \phi_{\vec k} / 2 }
\end{bmatrix} ,
\end{equation}
where
$ \cos \theta_{\vec k} = h_z (\vec k) /  [ h_x^2 (\vec k) + h_y^2 (\vec k) ]^{1/2}$ and $ \tan \phi_{\vec k} = h_y (\vec k) / h_x  (\vec k) $.
Such a system possesses PHS, and the particle-hole operation ${\cal P} = i \sigma_y {\cal K}$ (${\cal K}$ is the complex conjugation) transforms ${\cal P} | u_{\rm v, c} (\vec p ) \ra = i \tau_y {\cal K} | u_{\rm v, c} (\vec p ) \ra = |  u_{\rm c, v} (\vec p ) \ra$, i.e., each eigenstate $  | u_{\rm v, c} (\vec p ) \ra $ at energy $\epsilon_{\rm v, c} (\vec k)$ has a copy ${\cal P} | u_{\rm v, c} (\vec p ) \ra =  | u_{\rm c, v} (\vec p ) \ra $ at energy $\epsilon_{\rm c, v} (\vec k)$. Meanwhile it also satisfies ${\cal P} h (\vec k) {\cal P}^{-1} = - h (\vec k)$. Therefore $h (\vec k)$ has a particle-hole symmetry.

Due to the PHS possessed between the Bloch states $ |  u_{\rm c, v} (\vec p ) \ra = i \tau_y {\cal K} | u_{\rm v, c} (\vec p ) \ra $, we arrive at the following relation
\begin{align}
\vec r (\vec p, - \vec k) & = {\vec A}_{\rm c} (\vec p - \vec k/2 ) - {\vec A}_{\rm v} (\vec p + \vec k/2) - \nabla_{\vec p} \arg [ \la u_{\rm c} (\vec p - \vec k/2 ) | \hat{\boldsymbol \nu} | u_{\rm v} (\vec p + \vec k/2 ) \ra ] 
\nn
& = - {\vec A}_{\rm v} (\vec p - \vec k/2 ) + {\vec A}_{\rm c} (\vec p + \vec k/2) + \nabla_{\vec p} \arg [ \la u_{\rm v} (\vec p - \vec k/2 ) | \hat{\boldsymbol \nu} | u_{\rm c} (\vec p + \vec k/2 ) \ra^* ]
= \vec r (\vec p, \vec k) ,
\end{align}
which shows that $\vec r (\vec p,  \vec k) = \vec r (\vec p, - \vec k)$ is even with respect to $\vec k$.

\bigskip

\subsection{Derivation of $d_\alpha^\rho (\vec p)$ in the main text}

Our goal here is to calculate $d_\alpha^\rho (\vec p) =[ \p \rho (\vec p, \vec k) / \p k_\alpha]_{\vec k = \vec 0}  $, where
\be
\rho (\vec p, \vec k) = [ f (\epsilon_{ {\rm v}, \vec p - \vec k/2 }) - f (\epsilon_{ {\rm c}, \vec p + \vec k/2 }) ] \delta ( \omega_{ {\rm c}, \vec p + \vec k/2 } - \omega_{ {\rm v}, \vec p-\vec k/2 }  - \omega ) .
\ee
Therefore we expand $\rho (\vec p, \vec k)$ in terms of $k_\alpha$.
The first part of $\rho (\vec p, \vec k)$ can expanded out as
\be
[ f (\epsilon_{ {\rm v}, \vec p - \vec k/2 }) - f (\epsilon_{ {\rm c}, \vec p + \vec k/2 }) ] 
= [ f (\epsilon_{ {\rm v}, \vec p }) - f (\epsilon_{ {\rm c}, \vec p  }) ]
- \frac{k_\alpha}{2}  \frac{\p ( f_{ {\rm v}, \vec p } + f_{ {\rm c}, \vec p } ) }{ \p  p_\alpha }  + O(k^2) .
\label{eq:expandfermi}
\ee
To expand the second part of $\rho (\vec p, \vec k)$, i.e., $\delta ( \omega_{ {\rm c}, \vec p + \vec k/2 } - \omega_{ {\rm v}, \vec p-\vec k/2 }  - \omega )$, we introduce an auxiliary symmetrization:
\be
\omega_{ {\rm c}, \vec p}  = \omega_{\vec p}^{0}  + \tilde{\omega}_{{\rm c }, \vec p},
\quad
\omega_{ {\rm v}, \vec p}  = \omega_{\vec p}^{0}  + \tilde{\omega}_{{\rm v }, \vec p},
\ee
where $ \omega_{\vec p}^{0} = ( \omega_{{\rm c }, \vec p} + \omega_{{\rm v }, \vec p} ) / 2 $ is the the shared kinetic part, while $\tilde{\omega}_{{\rm c}, \vec p} =  ( \omega_{{\rm c }, \vec p} - \omega_{{\rm v }, \vec p} ) / 2 $ and $\tilde{\omega}_{{\rm v}, \vec p} =  - ( \omega_{{\rm c }, \vec p} - \omega_{{\rm v }, \vec p} ) / 2 $ are symmetrized conduction and valence band dispersions.
Using this auxiliary symmetrization, we have 
\be
\delta ( \omega_{ {\rm c}, \vec p + \vec k/2 } - \omega_{ {\rm v}, \vec p-\vec k/2 }  - \omega ) = \delta ( \tilde{\omega}_{ {\rm c}, \vec p + \vec k/2 } - \tilde{\omega}_{ {\rm v}, \vec p-\vec k/2 }  - \omega ) = \delta ( \omega_{\vec p}^{\rm cv} - \omega ) + O (k^2),
\quad
\omega_{\vec p}^{\rm cv} = \omega_{{\rm c }, \vec p}  - \omega_{{\rm v }, \vec p} ,
\label{eq:expanddelta}
\ee
where we used $\tilde{\omega}_{{\rm c }, \vec p} = - \tilde{\omega}_{{\rm v }, \vec p}$ in the second equation.
Combining the terms in Eqs.~(\ref{eq:expandfermi}) and (\ref{eq:expanddelta}), up to the first order of $k_\alpha$, we have 
\be
\rho (\vec p, \vec k) = \rho (\vec p, \vec 0) + k_\alpha^{} d_\alpha^\rho (\vec p) + O (k^2) ,
\ee
where
\be
\rho (\vec p, \vec 0) = [ f (\epsilon_{ {\rm v}, \vec p - \vec k/2 }) - f (\epsilon_{ {\rm c}, \vec p + \vec k/2 }) ]  \delta ( \omega_{\vec p}^{\rm cv} - \omega ) ,
\ee
that gives a vanishing shift current in a centosymmetric crystal,
and
\be
d_\alpha^\rho (\vec p)  = - \frac{1}{2} \frac{\p ( f_{ {\rm v}, \vec p } + f_{ {\rm c}, \vec p } ) }{ \p  p_\alpha }  \,  \delta ( \omega_{\vec p}^{\rm cv} - \omega ) ,
\ee
which is Eq.~(\ref{eq:d-alpha-rho}) in the main text.

\clearpage
\end{document}